\documentclass[%
 reprint,
 amsmath,amssymb,
 aps,
]{revtex4-2}

\usepackage{graphicx}
\usepackage{dcolumn}
\usepackage{bm}

\begin{document}

\title{Dynamics of electronic states in the insulating Intermediate surface phase of 1T-TaS$_2$}

\author{Jingwei Dong$^{1}$, Weiyan Qi$^{1}$, Dongbin Shin$^{2,3}$, Laurent Cario$^{4}$, Zhesheng Chen$^{5}$, Romain Grasset$^{1}$, Davide Boschetto$^{6}$, Mateusz Weis$^{6,7}$, Pierrick Lample$^{6,7}$, Ernest Pastor$^{8}$, Tobias Ritschel$^{9}$, Marino Marsi$^{5}$, Amina Taleb$^{10}$, Noejung Park$^{11}$, Angel Rubio$^{3}$, Evangelos Papalazarou$^{5}$ and Luca Perfetti$^{1}$}
\affiliation
{$^{1}$ Laboratoire des Solides Irradi\'{e}s, CEA/DRF/lRAMIS, Ecole Polytechnique, CNRS, Institut Polytechnique de Paris, F-91128 Palaiseau, France}
\affiliation
{$^{2}$ Department of Physics and Photon Science, Gwangju Institute of Science and Technology (GIST), Gwangju 61005, Republic of Korea }
\affiliation
{$^{3}$ Max Planck Institute for the Structure and Dynamics of Matter and Center for Free-Electron Laser Science, Luruper Chaussee 149, 22761 Hamburg, Germany}
\affiliation
{$^{4}$ Nantes Universit\'e, CNRS, Institut des Mat\'eriaux de Nantes Jean Rouxel, IMN, F-44000 Nantes, France}
\affiliation
{$^{5}$ Universit\'e Paris-Saclay, CNRS, Laboratoire de Physique des Solides, 91405 Orsay Cedex, France}
\affiliation
{$^{6}$ LOA, Laboratoire d'Optique Appliqu\'ee, CNRS, Ecole Polytechnique, ENSTA Paris, Institut
Polytechnique de Paris, 181 Chemin de la Hunière et des Joncherettes, 91120 Palaiseau, France.}
\affiliation
{$^{11}$ Universit\'e Paris-Saclay, CEA, CNRS, LIDYL, Gif-sur-Yvette, 91191, France}
\affiliation
{$^{8}$ IPR–Institut de Physique de Rennes, CNRS-Centre national de la recherche scientifique, UMR 6251 Université de Rennes, 35000 Rennes, France
}
\affiliation
{$^{9}$Instut f\"ur Festk\"orper- und Materialphysik, Technische Universit\"at Dresden, 01069, Dresden, Germany}
\affiliation
{$^{10}$ Synchrotron SOLEIL, Saint Aubin BP 48, Gif-sur-Yvette F-91192, France}
\affiliation
{$^{11}$ Department of Physics, Ulsan National Institute of Science and Technology (UNIST), UNIST-gil 50, Ulsan 44919, Korea}

\begin{abstract}

This article reports a comparative study of bulk and surface properties in the transition metal dichalcogenide 1T-TaS$_2$. When heating the sample, the surface displays an intermediate insulating phase that persists for $\sim 10$ K on top of a metallic bulk. The weaker screening of Coulomb repulsion and stiffer Charge Density Wave (CDW) explain such resilience of a correlated insulator in the topmost layers. Both time resolved ARPES and transient reflectivity are employed to investigate the dynamics of electrons and CDW collective motion. It follows that the amplitude mode is always stiffer at the surface and displays variable coupling to the Mott-Peierls band, stronger in the low temperature phase and weaker in the intermediate one.

\end{abstract}

\pacs{}

\maketitle

\section{introduction}

The transition metal 1T-TaS$_2$ crystallize in a trigonal antiprismatic structure with lattice constants $ a=3.36$ \AA~ and $c=5.9$ \AA. A pronounced Charge Density Wave (CDW) leads to a $\sqrt{13}\times\sqrt{13}$ superstructure \cite{Smaalen}, whose building blocks are clusters of 13 Ta atoms with the shape of David’s stars. 1T-TaS$_2$ has a rich phase diagram, ruled by the ordering of the CDW in a manifold of different structures with nearly equal free energy \cite{Wilson}. The superlattice pattern is Nearly Commensurate ($NC$) at room temperature and locks-in to a Commensurate ($C$) phase when cooling the sample below 174 K. During the heating cycle, a phase transition from the $C$ phase to a striped $T$ phase occurs at 223 K, while the $NC$ phase is recovered above 280 K. The $NC$ and $T$ phase are metallic, whereas the $C$ phase is insulating. These entwined orders emerge from the interplay of electron-phonon and electron-electron interactions, both being particularly strong in this dichalcogenide \cite{Fazekas,Perfetti2005,Papalazarou2023}. 

During the last 20 years, 1T-TaS$_2$ has been investigated by many experimental techniques, some being sensitive to the bulk, as transport, optical reflectivity and X-Ray Diffraction (XRD) measurements, whereas other being sensitive to the surface, as Angle Resolved Photoelectron Spectroscopy (ARPES) and Scanning Tunneling Microscopy (STM). Since the highest occupied band of monolayer 1T-TaS$_2$ is half filled, the insulating phase observed by STM arises from a Mott localization \cite{Monolayer}. Nonetheless, the interlayer coupling can strongly affect the electronic properties.

The David's stars of two adjacent layers can be: top Aligned ($A$), Laterally displaced with a vector of magnitude $2a$ ($L$) or Laterally displaced with a vector of magnitude $a$ ($L’$). X-Ray Diffraction (XRD) experiments have shown signature of $AL$ stacking in the $C$ phase \cite{Wang, Ritschel2015,Ritschel}. Scanning Tunneling Spectroscopy (STM) measurements show that the two topmost layers are often top aligned ($A$). Regardless of the dimers formation, strong electronic correlations open a Mott electronic gap also when the topmost layers are laterally displaced ($L$) \cite{Butler,Wu}. These conclusions have been also supported by ab-initio calculations of the electronic structure \cite{Werner,Jung}.

Despite the large amount of experiments that have been reported, only few works compared bulk and surface sensitive methods on the same batch of samples \cite{Wang,Burk}. Recently, Wang \textit{et al.} have shown that an intermediate phase ($I^s$) is observed by ARPES during the heating cycle, but cannot be detected by transport and XRD measurements\cite{Wang}. It follows that surface and bulk phase diagram of 1T-TaS$_2$ are different. The purpose of this article is to explore this dichotomy in detail, by making use of two bulk sensitive techniques (resistivity and optical pump-probe), and a surface sensitive technique (time resolved ARPES). By these means we reconstruct an extended phase diagram and characterize the properties of the $I^s$ phase. The analysis of electronic spectra is guided by state of the art ab-initio calculations based on Density Functional Theory plus Generalized Orbital U (DFT + GOU) \cite{Papalazarou2023}. The $I^s$ phase is ascribed to insulating layers without a well defined out-of planing stacking order and piled on top of a metallic bulk. Further insights can be extracted from the dynamics of the electronic states. The temporal oscillations of photoelectron spectra and transient reflectivity provide the frequency of the CDW amplitude mode in the topmost layers and inside the crystal, respectively. By comparing the two different measurements, it is shown that the CDW is systematically stiffer at the surface than in the bulk.

\section{methods}

\begin{figure}[t]
\includegraphics[width=\columnwidth]{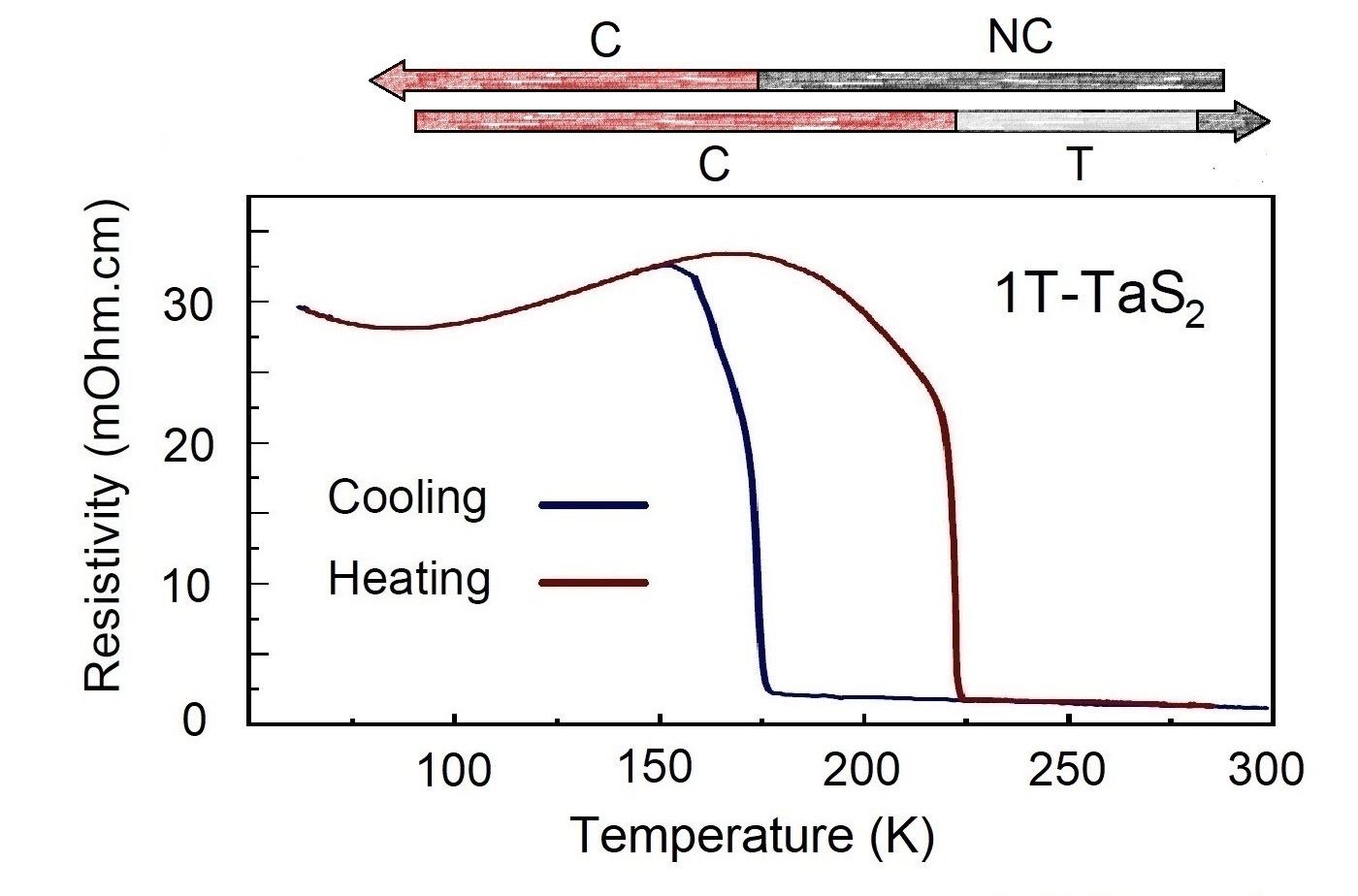}
\caption{Resistivity of 1T-TaS$_2$ during the cooling and heating cycle.}
\label{Temp}
\end{figure}

Single crystals of 1T-TaS$_2$ have been grown by vapor transport in the form of shiny plaquettes. The batch of samples has been characterized by x-ray diffraction and transport measurements. We used low-energy electron diffraction to verify the presence of the CDW reconstruction at the surface of the cleaved samples.

Photoelectron intensity maps have been acquired with $P$ polarization, photon energy of 6.2 eV and integrating the photoemission intensity in a wavevector interval of $\pm 0.5 $ nm$^{-1}$ around the center of the Brillouin Zone. The 1T-TaS$_2$ have been cleaved at 300 K and measured at base pressure below $1\times10^{-10}$ mbar. Since the estimated photoelectron escape depth of such low photon energy electrons is of 3-5 nm, the ARPES and time resolved ARPES data are sensitive to roughly 5-8 layers from the termination of the 1T-TaS$_2$ sample \cite{Chen}. In the rest of this work, we will refer to such region as the probed \lq\lq surface”. In the case of time resolved ARPES, the pump pulse is centered at 1.55 eV and has a fluence of 180 $\mu$J/cm$^2$. The cross correlation between pump and probe is $\approx 100$ fs \cite{Faure}.

\begin{figure}[t]
\includegraphics[width=\columnwidth]{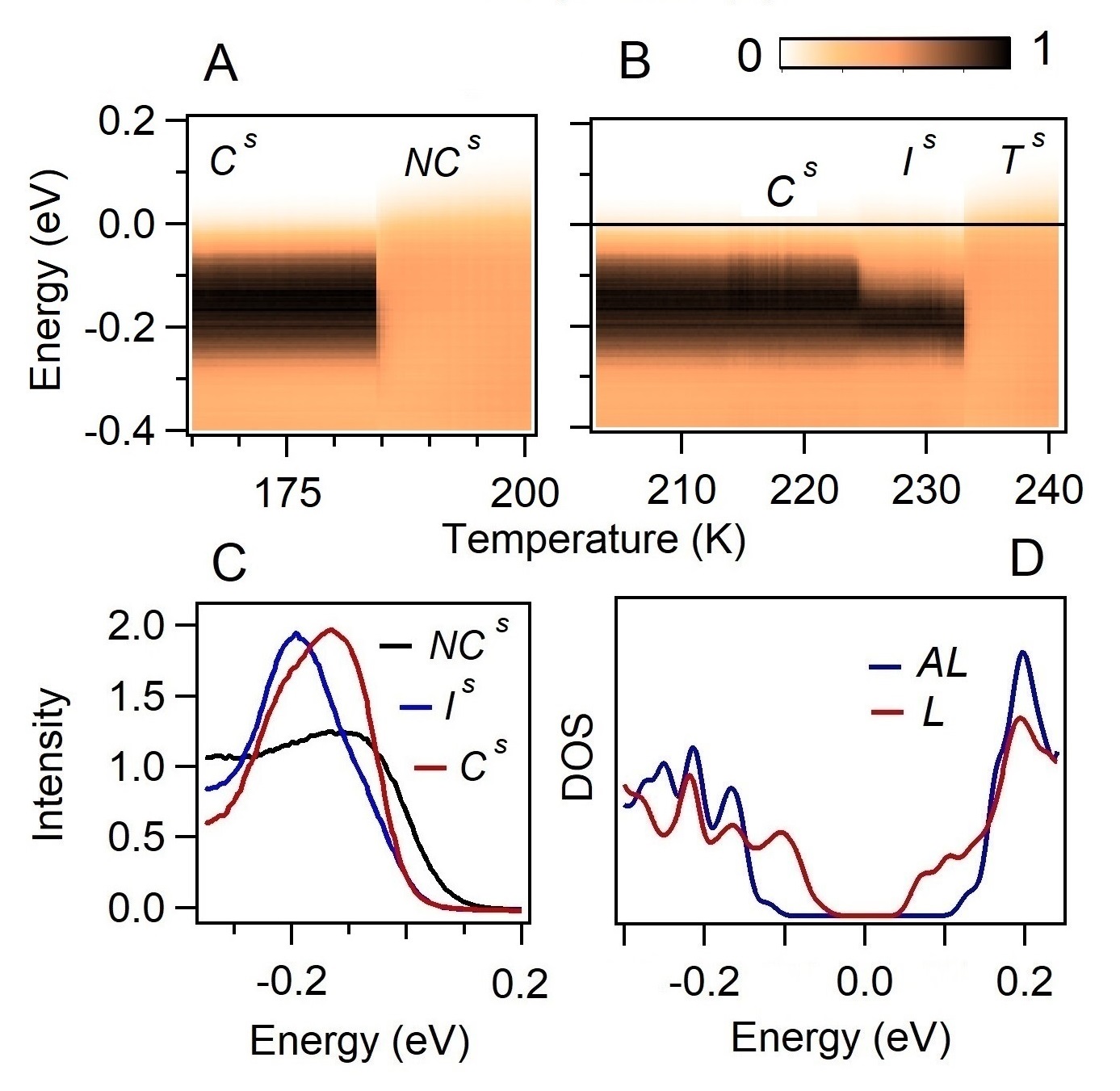}
\caption{A) Evolution of photoelectron intensity during the cooling cycle. B) Evolution of photoelectron intensity during the heating cycle. C) Energy Distribution Curves of the $NC^s$, $C^s$ and $I^s$ phase. D) Density Of electronic States (DOS) calculated by self-consistent DFT-GOU method for the $L$ and $AL$ stacking. The self-consistent $\overline U_{AL}=0.45$ eV and $\overline U_{L}=0.33$ eV are obtained via the ABCN0 method.}
\label{Temp}
\end{figure} 

Transient reflectivity measurements have been done with a pump beam centered at 1.55 eV and fluence of 180 $\mu$J/cm$^2$ (as for the time resolved ARPES). The probing beam is centered at 2100 nm (0.6 eV) and also gives a cross correlation of $\approx 100$ fs. The polarization of the probe is orthogonal to the one of the pump and the probe is in nearly normal incidence. We mounted the sample on an optical cryostat to perform temperature dependent measurements.

Density Functional Theory (DFT) calculations have been done by using the Quantum ESPRESSO package with PBE-type functional. Wave functions are obtained via the projector-augmented plane wave method and a basis set with a cutoff energy of 60 Ry. The lattice constant for relaxed bulk structure of 1T-TaS$_2$ are $a=3.36$ \AA, $c=6.03$ \AA~ and a $3 \times 3 \times 6$ k-point mesh samples the Brillouin zone. In the case of DFT + GOU, the $\overline U$ potential of a SD cluster is obtained self consistently via the ACBN0 method, where $\overline U$ is determined through the theory of screened Hartree-Fock exchange potential in a correlated subspace \cite{Papalazarou2023, Rubio}.

\begin{figure}[htp]
\includegraphics[width=\columnwidth]{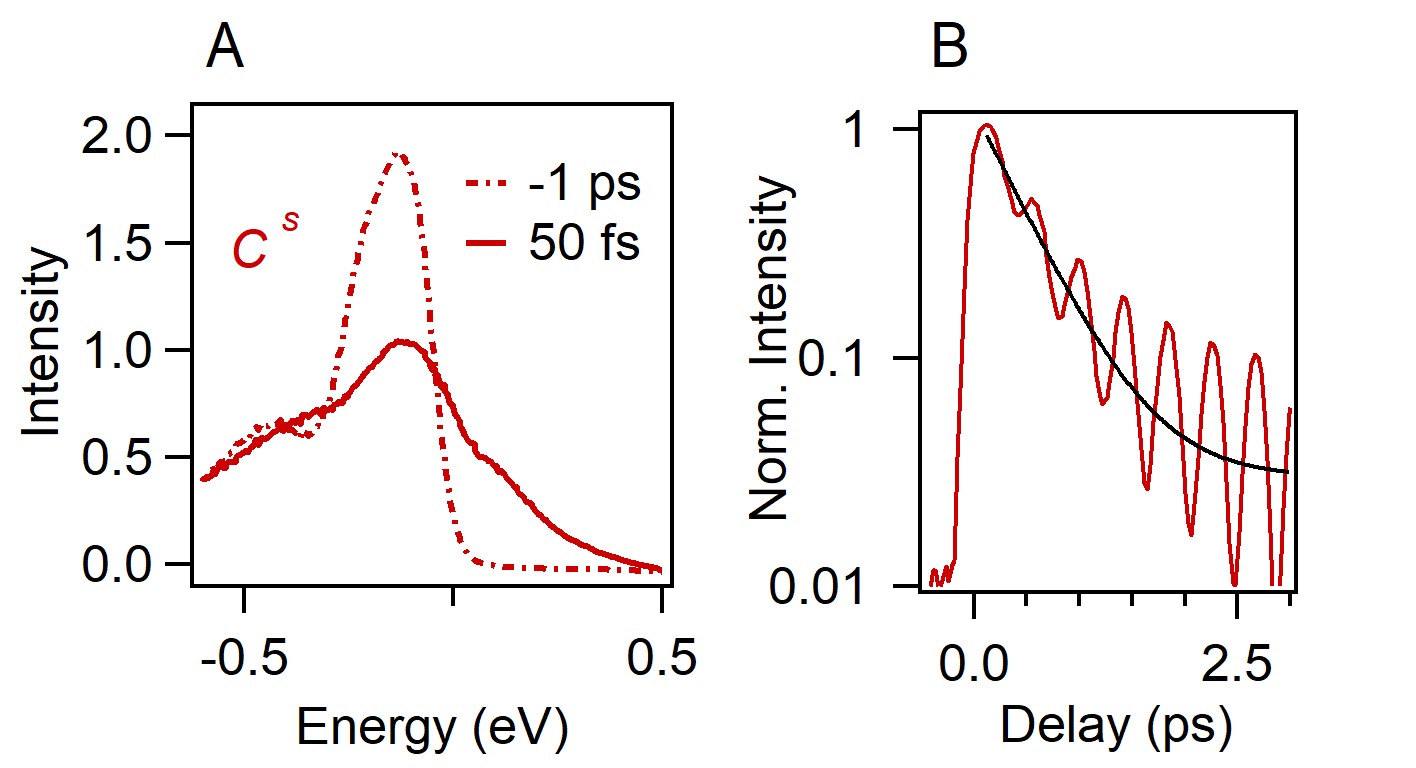}
\caption{ A) EDCs acquired in the $C^s$ phase at 140 K, at negative delay (dashed curve) and 50 fs after the arrival of the pump pulse (solid curve) . B) Photoelectron intensity of the $C^s$ phase integrated in the energy interval [0.05,0.4] eV as a function of pump probe delay (red curve) and exponential fit with decay time of 0.45 ps (black curve).}
\end{figure}

\begin{figure}[htp]
\includegraphics[width=\columnwidth]{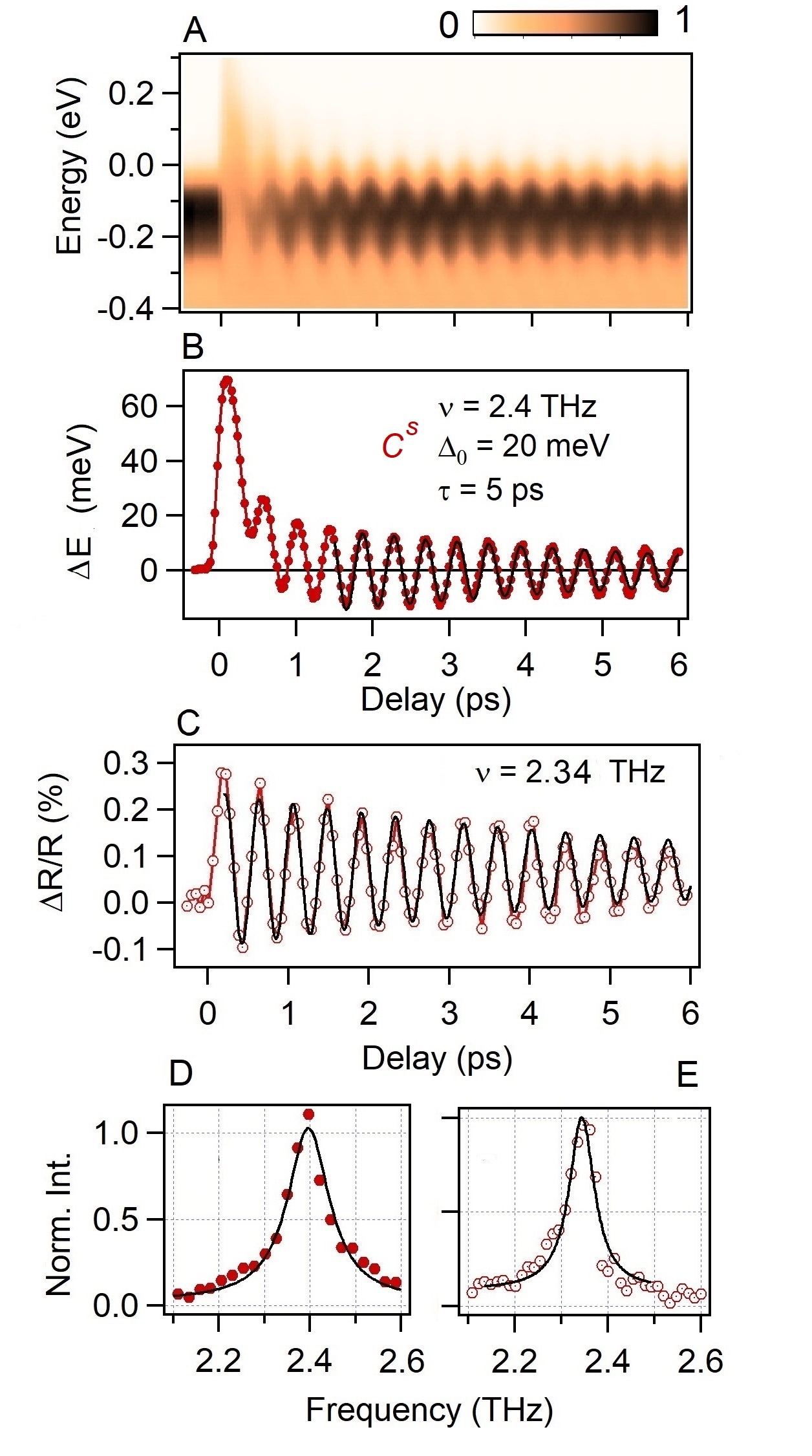}
\caption{ A) Photoelectron intensity of the photoexicted state at 140 K ($C^s$ phase). B) Mean energy shift extracted from the EDCs of the intensity map. C) Transient Reflectivity measured with probe frequency of 0.6 eV and 140 K ($C$ phase). D) Fourier transform of the CDW amplitude mode measured by time resolved photoemission at 140 K. E) Fourier transform of the CDW amplitude mode measured by transient reflectivity at 140 K.}
\label{CS}
\end{figure}

\section{Spectroscopic properties and surface phase diagram}

We show in Fig. 1 the resistivity of 1T- TaS$_2$ as a function of temperature. During the cooling cycle, the transition from the metallic $NC$ phase to the insulating $C$ phase takes place at 174 K. Upon heating the sample, the resistivity displays a transition from the $C$ to the metallic $T$ phase at 223 K. These data are comparable with previously published results \cite{Wang}, although the transition temperature may vary slightly from one batch to the other. 

Next we focus on the spectral properties of the surface. Figure 2A shows the photoelectron intensity map acquired during the cooling cycle. The transition from the $NC^s$ to the $C^s$ phase takes place when the temperature is lowered below 184 K (the suffix “s” in $NC^s$ and $C^s$ stand for “surface”). Figure 2B displays instead the intensity map acquired during a heating cycle. The CDW enter in to the intermediate insulating phase $I^s$ at 226 K, namely when the bulk resistivity drops to the metallic level. Roughly 10 K above this set point, the surface displays a transition to the metallic $T^s$ phase. These results are reproducible and have been repetitively observed by cycling the temperature more than once on different samples. Moreover, the present findings are in line with the data measured by Wang \textit{et al.} \cite{Wang} at photon energy of 21 eV and with Bao \textit{et al.} \cite{Zhou} at photon energy of 6.3 eV. The stability of the $I^s$ phase over a metallic bulk is ensured by the lower coordination number at the surface, which results in a stiffer CDW and larger Coulomb repulsion. The stiffening of David's stars is likely due to a relaxation of the lattice spacing in near surface layers, which can better accommodate the buckling of the sulfur atoms along the c-axis direction.
Larger correlation effects can be instead reproduced by performing self-consistent ABCN0 calculations of the screened Coulomb potential $\overline U$ in four layers 1T-TaS$_2$. The resulting $\overline U$ is roughly 20\% larger in the outer two layers than in the two inner ones. This property is common to many correlated materials. Note that an insulating surface, distinct from the bulk, has been also reported in the related compound 1T-TaSe$_2$ \cite{Perfetti2003,Perfetti2005b,Yin}. The latter compound has a metallic bulk down to low temperature, while the surface becomes insulating for $T< 250$ K.

\begin{figure}[htp]
\includegraphics[width=\columnwidth]{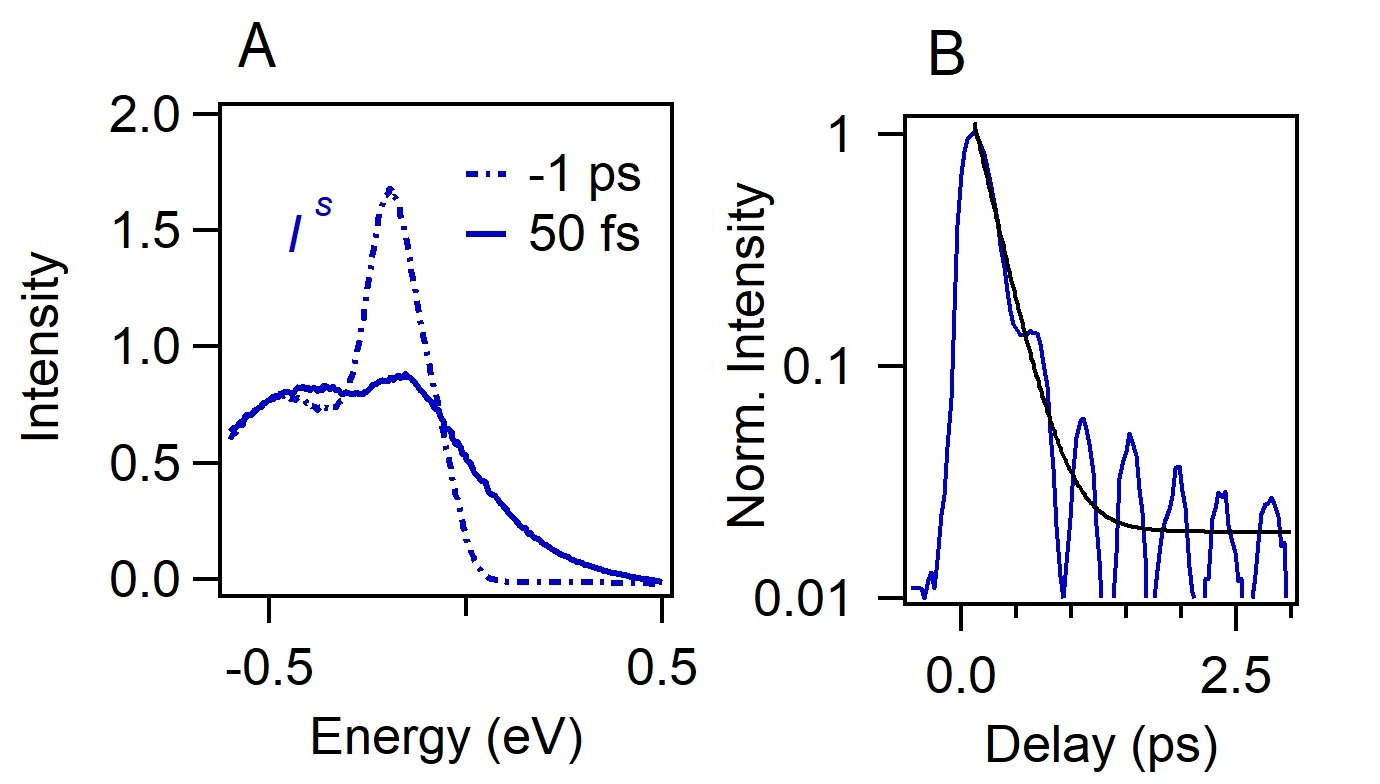}

\caption{ A) EDCs acquired in the $I^s$ phase at 220 K, at negative delay (dashed curve) and 50 fs after the arrival of the pump pulse (solid curve). B) Photoelectron intensity of the $I^s$ phase integrated in the energy interval [0.05,0.4] eV as a function of pump probe delay (blue curve) and exponential fit with decay time of 0.2 ps (black curve).}
\end{figure}

\begin{figure}
\includegraphics[width=\columnwidth]{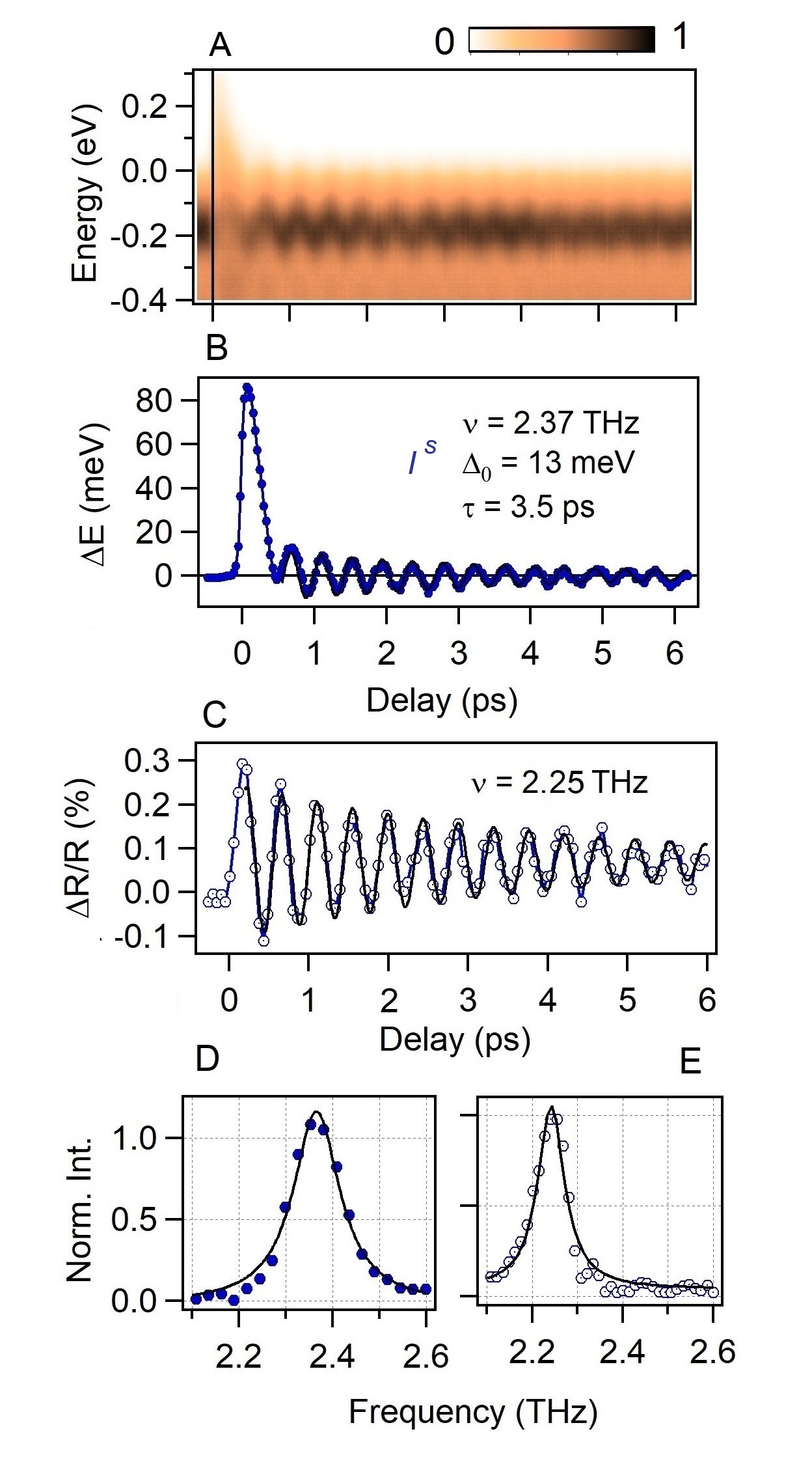}
\caption{ A) Photoelectron intensity of the photoexicted state at 225 K ($I^s$ phase). B) Mean energy shift extracted from the EDCs of the intensity map. C) Transient Reflectivity measured with probe frequency of 0.6 eV and 210 K ($C$ phase). D) Fourier transform of the CDW amplitude mode measured by time resolved photoemission at 225 K. E) Fourier transform of the CDW amplitude mode measured by transient reflectivity at 210 K.}
\label{IS}
\end{figure}

Figure 2C displays Energy Distribution Curves (EDCs) that have been extracted in the $NC^s$, $C^s$and $I^s$ phase. The metallic $NC^s$ holds a fermi level, whereas the $C^s$ and $I^s$ have vanishing density of the electronic states at the chemical potential. Note that the spectral properties of the two insulating states are qualitatively different: the EDC of the $C^s$ phase peaks at $-0.1$ eV while the EDC of the $I^s$ phase peaks at $-0.2$ eV. Wang et al. have shown that EDCs of the $C^s$ phase strongly disperses in the c-axis direction whereas the EDCs of the $I^s$  phase do not. The reduction of bandwidth can explain the different EDC of $I^s$ phase. Here, we propose that the lack of interlayer dispersion is due to the randomization of CDW stacking in the c-axis direction.

The vanishing spectral weight at the chemical potential (see Fig. 2C) indicates that the top 4-6 layers of the $I^s$ phase are mostly insulating. Indeed, the Coulomb repulsion among electrons should be even more effective when stacking disorder reduces the out-of-plane dispersion. Besides, the self-consistent DFT+GOU calculations in Fig. 2D reveal a gapped Density of Electronic States (DOS) for commensurate structures with different stacking configurations.

In contrast, the nearly commensurate $NC$ and $T$ phases hold metallic conductivity. As a consequence, we propose that the CDW of $I^s$ state gradually loses in-plane commensurability within the depth of the crystal, over a characteristic distance comparable or larger than the photoelectron mean free path. The crossover could be described by a variable filling factor of metallic domains \cite{Monney}, reconciling the insulating spectral properties near to the surface with the metallic conductivity of the bulk.

\section{Dynamics of electrons and CDW amplitude: time resolved ARPES vs transient reflectivity}

The dynamics of the CDW has been first reported by transient reflectivity experiments \cite{Demsar}. The breathing mode of David's stars clusters, also identified with the amplitude mode of the CDW, always dominates the time dependent response \cite{Mihailovic}. Moreover, sizable softening of the CDW amplitude mode has been observed when approaching the transition temperature from the insulating side \cite{Demsar}. Time resolved ARPES experiments have shown that the electrons near to the chemical potential and the electronic bandgap is strongly coupled to the CDW amplitude \cite{Perfetti2006,Perfetti2008}. The periodic oscillations of these states are superimposed to an ultrafast metal-insulator transition that fills the gap via strong CDW fluctuations \cite{Papalazarou2023}.

Here we investigate the dynamics of electrons and collective modes in the different phases at the surface. Figure 3A shows the EDCs before and just after photoexcitation in the $C^s$ phase at 140 K. Just after photoexcitation, the appearance of spectral weight at the chemical potential indicates an ultrafast melting of the correlated gap. Dong et al. have recently shown that CDW fluctuations induce a localization of low energy states during this ultrafast phase transition \cite{Papalazarou2023}. The characteristic cooling time of hot electrons is estimated by integrating the signal above the chemical potential in the interval [0.05, 0.4] eV. As shown in Fig. 3B the exponential fit gives a decay time of $\tau_e=0.45$ ps, which is in line with previously reported data \cite{Perfetti2006,Perfetti2008}.

As expected, the photoelectron intensity map in Fig. 4A show large oscillations of the Lower Peierls Mott Band (LPMB), lasting for tens of picoseconds. This effect is quantified in Fig. 4B, which plots the variation mean EDCs energy as a function of time. The latter is defined as:
$$ \Delta E (t) = \frac{ \int I(E,t) E dE}{\int I(E,t)  dE}, $$
where the integral is performed within the interval $[-0.4, 0.4]$ eV and $I(E,t)$ describes the EDC at pump probe delay $t$. We model the $\Delta E (t)$ oscillations (black solid line in Fig. 4B) by the function $\Delta E_0 \cos(2\pi \nu t)\exp(-t/\tau_d)$. The best fit is obtained with frequency $\nu=2.4$ THz, damping time of $\tau_d=5$ ps and $\Delta E_0=20$ meV. As a term of comparison, we also measured the photoinduced modulation \cite{Demsar} $\Delta R/R$, where $R$ is the reflectivity of probe beam. Fluence and center frequency of the pump pulses are the same as for the time resolved ARPES experiment while the probe beam has been centered at 2100 nm ( 0.59 eV). Being bulk sensitive, the transient reflectivity provides information about the bulk $C$ phase. Moreover we always compare data at equal (or nearby) temperature because the frequency of the amplitude mode strongly depends on this parameter \cite{Demsar}. By fitting $\Delta R/R$ with a damped cosinus (see Fig. 4C) we find $\nu=2.36$ THz and damping time of $\tau_d=6$ ps. Figure 4D,E display the Fourier transform of the oscillations observed in tr-ARPES and transient reflectivity, respectively. The amplitude mode is 60 GHz stiffer at the surface (Fig. 4D) than in the bulk (Fig. 4E). Namely, the lower coordination number favors a stiffer, and presumably larger, CDW amplitude.

Analogous measurements and analysis have been repeated at 225 K, when the surface is in the $I^s$ phase. As shown in Fig. 5A, also in this case the photo excitation induces an ultrafast melting of the correlated gap. However, the dynamics of integrated signal in Fig. 5B indicates that cooling of hot electrons takes place on a timescale $\tau_e=0.2$ ps, which is twice faster than the cooling observed in the $C^s$ phase. Similar observations have been recently reported in another tr-ARPES investigation of the $I^s$ phase and have been ascribed to strong electron-electron correlations \cite{Zhou}. Instead, we propose that the fast energy dissipation of the hot electrons in the $I^s$ phase is due to the presence of a nearby metallic bulk and metallic surface domains \cite{Monney,Yin}, in which phonons can be absorbed very effectively because of the gapless electronic DOS. 
Figure 6A shows the photoelectron intensity map of the $I^s$ phase as a function of pump-probe delay. The $\Delta E (t)$ evolution is plot in Fig. 6B. In the case of $I^s$ surface, the best fit to the $\Delta E$ oscillations lead to a bare frequency $\nu=2.37$ THz, damping time $\tau_d=3.2$ ps and $\Delta E_0=13$ meV (see Fig. 6B). The transient reflectivity measurements at 210 K (see Fig. 6C) give instead $\nu=2.25$ THz and damping time $\tau_d=3$ ps. Again, the amplitude mode is stiffer at the surface than in the bulk. By comparing the Fourier transform of the oscillations observed by the two measurements (Fig. 6D vs Fig. 6E) we can visually appreciate the significance of this difference: the CDW has a frequency 120 GHz larger in the $I^s$ phase (Fig. 6D) than the $C$ phase at nearby temperature (Fig. 6E).

Finally, we compare the amplitude of oscillations that we have observed in the different phases. Note that $\Delta E (t)$ oscillations are 40\% smaller in the $I^s$ (225 K) than in the $C^s$ phase (140 K) (see Fig. 4B and 6B). Instead, the transient reflectivity shows that the initial displacement of CDW mode is almost equal in the $C$ phase at 210 K and 140 K (see Fig. 4C and 6C). We deduce that the coupling of the low energy electronic states to the CDW amplitude is considerably smaller in the $I^s$ phase than in the $C^s$ one. Eventually, the disordered staking may cause weaker electron phonon coupling strength. In this context it is worth mentioning that STM has shown that electron-phonon coupling depends on the surface termination. High resolution measurements resolved narrow emission lines arising from inelastic scattering of electrons with the amplitude mode \cite{Iwasa}. The progression of Franck-Condon replicas, and therefore the electron-phonon interaction strength, turns out to be dependent on the stacking configuration.

\section{Conclusions}

In conclusion we show that bulk and surface of 1T-TaS$_2$ display different electronic properties. The intermediate phase taking place during the heating cycle is ascribed to insulating surface layers that lost stacking order. The reduced screening of Coulomb repulsion and lack of interlayer dispersion can boost the correlation effects in the $I^s$ phase, making the near surface layers insulating even if the bulk is metallic. As for the electronic properties, also the CDW stiffness depends on the depth from the crystal termination. We observed different dynamics of CDW mode at the surface and in the bulk, both in the $C^s$ and $I^s$ phase. The amplitude mode in the $C^s$ phase is 60 GHz stiffer than one of the $C$ phase at equal temperature. This dichotomy doubles in the case of the intermediate phase, since the amplitude mode in the $I^s$ phase is 120 GHz stiffer than in the $C$ phase at a nearby temperature. Our findings could be general to other correlated CDW systems, since the lower coordination of the topmost layer should naturally enhance correlation effects and ease the structural distortion induced by the CDW.

\section{Acknowledgments}

R. G. acknowledges the financial support by “Investissements d’Avenir” LabEx PALM (ANR-10-LABX-0039-PALM) and T. R.  acknowledges financial support by the Würzburg-Dresden Cluster of Excellence on Complexity and Topology in Quantum Matter-ct.qmat (EXC 2147, project-id 390858490).
Jingwei Dong and Weiyan Qi have equally contributed to this article.

\section{Appendix A}

Figure 7 shows $\Delta E$ of the $C^s$ phase on a time interval of 20 ps. (Namely the same data of Fig. 4B but plot on a longer delay scan). There is no sign of the beating reported by Perfetti \textit{et al.} in the first tr-ARPES measurements of 1T-TaS$_2$ \cite{Perfetti2006}. We believe that a higher concentration of defects in our samples reduces by a factor two the decay time of the CDW amplitude mode.  In this case, no oscillation can be observed after 10 ps, so that the beating due to the weaker oscillatory component is no longer detectable.

\begin{figure}[htp]
\includegraphics[width=\columnwidth]{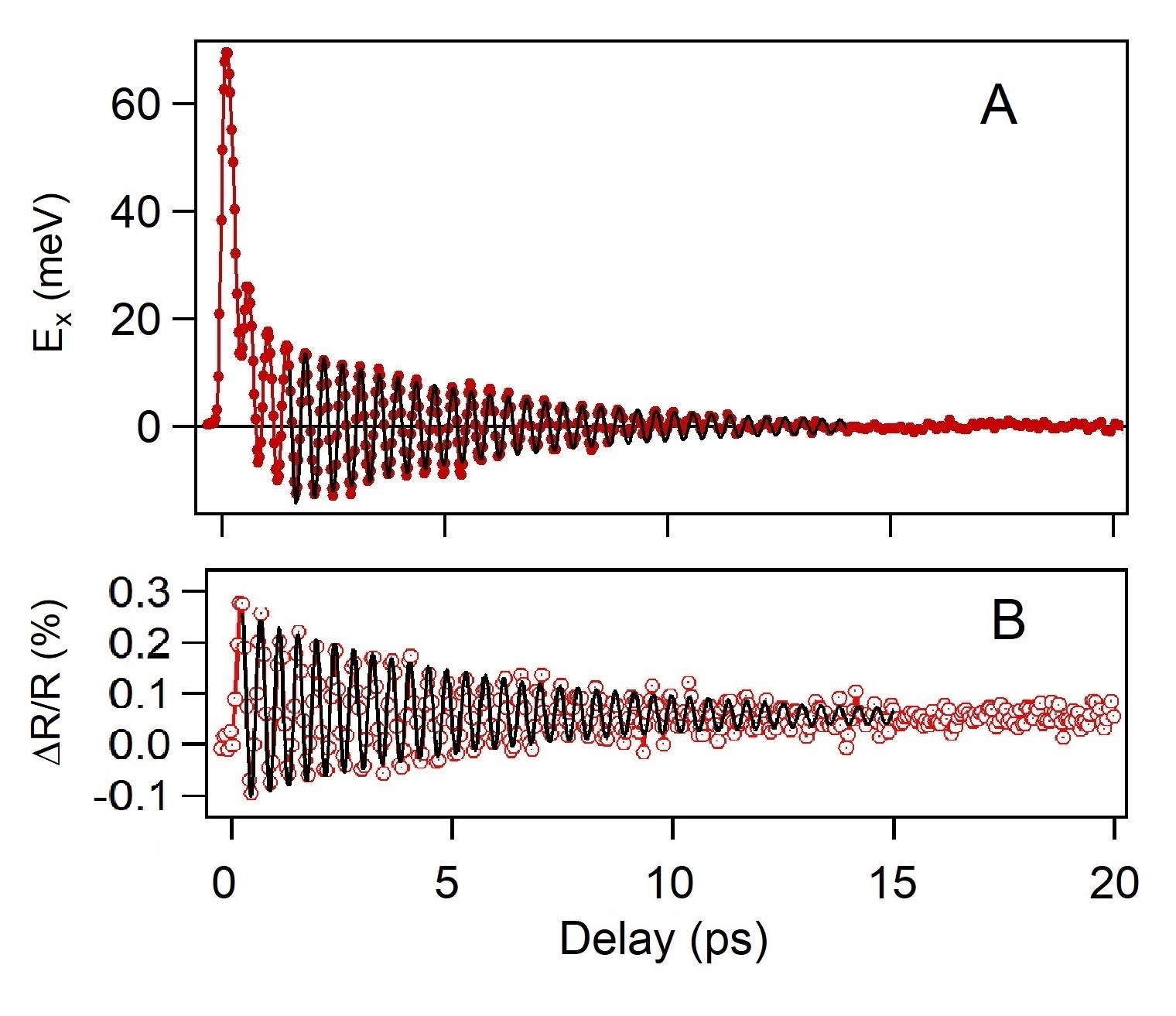}
\caption{A) Evolution of mean energy shift extracted from the photoectron signal acquired in the $C^s$ phase at 140 K. B) Transient Reflectivity measured with probe frequency of 0.6 eV and 140 K ($C$ phase).}
\end{figure}

\end{document}